\documentclass[journal]{IEEEtran}
\IEEEoverridecommandlockouts
\usepackage{cite}
\usepackage{overpic}
\usepackage{amsmath,amssymb,amsfonts}
\usepackage{graphicx}
\usepackage{textcomp}
\usepackage{xcolor}
\usepackage{float}
\usepackage{amsthm}
\usepackage{graphicx}
\usepackage{epstopdf}
\usepackage{amsmath,bm}
\usepackage{amsfonts}
\usepackage{amssymb}
\usepackage{color}
\usepackage{subfigure}
\usepackage{multirow}
\usepackage{multicol}
\usepackage{soul,xcolor}
\usepackage{algorithm}
\usepackage{algpseudocode}%

\theoremstyle{plain}

\newcommand{\vect}[1]{\mathbf{#1}}

\def\tr{\mathrm{tr}}

\def\Htran{\mbox{\tiny $\mathrm{H}$}}
\def\Ttran{\mbox{\tiny $\mathrm{T}$}}

\begin{document}

\title{Multi-Target Integrated Sensing and Communications in Massive MIMO Systems \vspace{-4mm}
}

\author{Ozan Alp Topal, Özlem Tuğfe Demir, Emil Björnson, and Cicek Cavdar \vspace{-7mm}

\thanks{O. A. Topal, E. Björnson, and  C. Cavdar are with the School of Electrical Engineering and Computer Science, KTH Royal Institute of Technology, Stockholm, Sweden (e-mail: \{oatopal, emilbjo, cavdar\}@kth.se). Ö. T. Demir is  with the Department of Electrical and Electronics Engineering, TOBB University of Economics and Technology, Ankara, Türkiye, (e-mail: ozlemtugfedemir@etu.edu.tr).}  
\thanks{This work has been funded by Celtic-Next project RAI-6Green partly supported by Swedish funding agency Vinnova and SSF SUCCESS project.
}
}

\maketitle

\begin{abstract}
Integrated sensing and communications (ISAC) allows networks to perform sensing alongside data transmission. While most ISAC studies focus on single-target, multi-user scenarios, multi-target sensing is scarcely researched. This letter examines the monostatic sensing performance of a multi-target massive MIMO system, aiming to minimize the sum of Cram\'{e}r-Rao lower bounds (CRLBs) for target direction-of-arrival estimates while meeting user equipment (UE) rate requirements. We propose several precoding schemes, comparing sensing performance and complexity, and find that sensing-focused precoding with power allocation for communication achieves near-optimal performance with 20 times less complexity than joint precoding. Additionally, time-sharing between communication and sensing outperforms simple time division, highlighting the benefits of resource-sharing for ISAC.
\end{abstract}

\vspace{-2mm}

\vspace{-4mm}
\section{Introduction}
\vspace{-1mm}
Integrated sensing and communications (ISAC) is anticipated to be a key feature of 6G and beyond networks, enabled by advancements in sensing and communication radios with larger antenna arrays and higher frequencies. The integration enhances communication efficiency through better environmental awareness and opens new revenue streams for operators by offering new sensing applications. A notable example is localization for autonomous vehicles and drones, where 6G is expected to deliver cm-level accuracy for customers \cite{ISAC_main}.

Direction-of-arrival (DoA) estimation is the final step of the localization of a target. Multiple-input multiple-output (MIMO) radar systems with sufficiently many antennas can provide less than one degree of estimation error thanks to the beamforming capabilities \cite{bekkerman2006target}. For precise localization, communication networks must offer similar sensing accuracy. Cram\'{e}r-Rao lower bound (CRLB) is commonly used to estimate the upper limit of sensing performance in target localization studies \cite{isac_metric}. The radar sensing capability of a  MIMO communication network is shown to provide limited degrees-of-freedom (DoF) without designing any additional sensing signals \cite{DoF_sensing}. In \cite{isac_CRB_main}, a joint design of sensing and communication (S\&C) precoding is proposed to minimize the CRLB on the DoA of a single target. The joint S\&C precoding is demonstrated to be energy efficient in \cite{energy_efficient}. 
However, these works consider only the existence of a single target, while in a practical scenario, multiple targets are expected to be sensed by the communication network simultaneously.

Multi-target sensing has been researched from the radar perspective without the consideration of the communication integration\cite{bekkerman2006target,range_sensing}. Only a handful of papers are available for the multi-target, multi-user MIMO scenario \cite{multicast_multitarget, RIS_multitarget, zhu2023information}, where user equipment (UE) interference and separate sensing signal design are not considered. Adding more targets complicates the trade-off, as precoders must balance communication and sensing with shared resources, while various configurations create intricate interference patterns.
Most closely related to our work, an optimal joint S\&C precoding scheme is proposed for multi-user multi-target bi-static sensing with the objective of weighted sum-CRLB minimization under the UEs' sum-rate constraints in \cite{zhu2023information}. However, their solution cannot be applied when UEs have individual rate constraints because their semi-definite relaxation (SDR)-based joint S\&C precoding is expected to no longer guarantee a rank-one solution due to the high number of constraints \cite{SDR_high_constraint}. Furthermore, none of these works has analyzed the potential gains or losses of transmitting sensing and communication signals within the same time/frequency resource block. 

In this letter, we study the performance of monostatic sensing for multi-target in the downlink of a massive MIMO system \cite{massivemimobook}. We aim to minimize the estimation error regarding DoA while guaranteeing individual rate constraints of UEs. We propose several precoding schemes and compare the sensing performance and implementation complexity. Our numerical analysis illustrates that joint precoding design for sensing and communication is unnecessary. Instead, choosing high-performance interference cancellation-capable precoding vectors for the communication signals, and optimizing the precoding only for the sensing signal performs similarly to the joint design and reduces the complexity by $95\%$. The optimal precoder lies predominantly in the nullspace of the UE channels, and in the derivative space of the target channels, demonstrating an untypical structure of the optimal precoding due to multi-targets. Furthermore, we compare the cases when the time duration is divided between sensing and communication and when the time is shared. We observe that the time-sharing case performs better than dividing the time duration. In the time-sharing case, communication signals help the radar sensing especially when targets and UEs are co-located, demonstrating the potential of resource sharing for ISAC in massive MIMO systems.

\vspace{-4mm}

\section{System Model}

We consider a downlink massive MIMO ISAC system consisting of a dual-functioning base station (BS) equipped with $M$ transmit and $M$ receive antennas, $K$ single-antenna communication UEs, and $T$ sensing target locations with the corresponding DoAs $\varphi_1,\ldots,\varphi_T$. From the transmit antennas, at channel use $n$, the following signal is transmitted:
\vspace{-2mm}
\begin{equation}
    \vect{x}[n] =\sum_{k=1}^{K}\vect{w}_ks_k[n] + \sum_{t=1}^{{\bar{T}}}\bar{\vect{w}}_t\bar{s}_t[n] , \quad n=1,\ldots,N, \vspace{-2mm} 
\end{equation}
where $\vect{w}_k \in \mathbb{C}^M$ and $\bar{\vect{w}}_t \in \mathbb{C}^M$ are the precoding vector for UE $k$ and for sensing stream $t$, respectively. Here, $\bar{T}$ is the number of sensing streams, which might be different than $T$. The total number of channel uses is $N$. The power allocated to the UEs and targets can be given as $\|\vect{w}_k\|^2 = p_k$ and $\|\bar{\vect{w}}_t\|^2 = \bar{p}_t$, respectively, where the total transmit power is $\sum_{k=1}^{K}p_k+\sum_{t=1}^{{\bar{T}}}\bar{p}_t = P_{\textrm{max}}$. $s_k[n]$ is the zero-mean unit-power information symbol of UE $k$, and $\bar{s}_t[n]$ is the zero-mean unit-power sensing symbol of the stream $t$. The information symbols are mutually independent of each other and the target symbols. The received signal at UE $k$  is written as
\vspace{-2mm}
\begin{equation}
    y_k[n]=\vect{h}_k^{\Ttran}\vect{x}[n]+\zeta_k[n], \vspace{-2mm}
\end{equation}
where  $\zeta_k[n] \sim  \mathcal{N}_{\mathbb{C}}(0,\sigma^2)$ is the additive white Gaussian noise. $ \mathbf{h}_{k} \in \mathbb{C}^{M}$ denotes the channel from the BS to the UE $k$, which is perfectly known at the BS.\footnote{
Perfect CSI is assumed to concentrate on the best achievable performance of multi-target massive MIMO system.} The signal-to-interference-plus-noise ratio (SINR) of UE $k$ is
\vspace{-2mm}
\begin{equation}
    \mathrm{SINR}_k = \frac{|\vect{h}_k^{\Ttran}\vect{w}_k|^2}{\sum_{i=1,i\neq k}^K|\vect{h}_k^{\Ttran}\vect{w}_i|^2 + \sum_{t=1}^{{\bar{T}}} |\vect{h}_k^{\Ttran}\bar{\vect{w}}_t|^2 + \sigma^2}, \vspace{-2mm}
\end{equation}
and the achievable rate of UE $k$ is $\mathrm{R}_k = B\log_2(1+\mathrm{SINR}_k)$, where $B$ denotes the operational bandwidth.

For the sensing part, the BS utilizes echo signals collected during $N$ channel uses to carry out DoA estimation of the $T$ targets. 
The sensing received signal is
\vspace{-2mm}
\begin{equation}
    \vect{y}[n] = \sum_{t=1}^T \alpha_t\underset{=\mathcal{A}(\varphi_t) }{\underbrace{ \vect{a}(\varphi_t)\vect{a}^{\Ttran}(\varphi_t)}}\vect{x}[n]+\boldsymbol{\zeta}[n],
    \label{eq:target_received_signal} \vspace{-2mm}
\end{equation}
where $\alpha_t\in \mathbb{C}$ is the radar cross section (RCS) of the target $t$, which also includes the effect of round-trip path loss of the mono-static channel, and $\vect{a}(\varphi_t) \in \mathbb{C}^M$ is the array response vector. The entries of the noise $\boldsymbol{\zeta}[n]$ are independent and identically distributed with $\mathcal{N}_{\mathbb{C}}(0,\sigma^2)$. Note that the targets are assumed to be in the same Doppler-range bin.\footnote{When each target is located in a distinct Doppler-range bin, \eqref{eq:target_received_signal} can instead be simplified for each target by setting $T=1$.} We can concatenate the unknown but deterministic parameters to be estimated as $\boldsymbol{\xi} = \left[ \boldsymbol{\varphi}^{\Ttran}, \boldsymbol{\alpha}^{\Ttran}  \right]^{\Ttran}$, where  $\boldsymbol{\varphi} = [\varphi_1,\ldots,\varphi_T]^{\Ttran}$, and  $\boldsymbol{\alpha} = [\operatorname{Re}\left(\alpha_1\right),\operatorname{Im}\left(\alpha_1\right),\ldots,\operatorname{Re}\left(\alpha_T\right),\operatorname{Im}\left(\alpha_T\right)]^{\Ttran}$. The Fisher information matrix for estimating $\boldsymbol{\xi}$ is given from
\cite{bekkerman2006target} 
as
\vspace{-2mm}
\begin{equation}
    \vect{J} = \begin{bmatrix} \vect{J}_{\boldsymbol{\varphi}\boldsymbol{\varphi}} &\vect{J}_{\boldsymbol{\varphi}\boldsymbol{\alpha}} \\ \vect{J}_{\boldsymbol{\varphi}\boldsymbol{\alpha}}^{\Ttran} & \vect{J}_{\boldsymbol{\alpha}\boldsymbol{\alpha}} \end{bmatrix}. \vspace{-2mm}
\end{equation}
The elements of the Fisher matrix are obtained as in \cite{bekkerman2006target}
\vspace{-2mm}
\begin{equation}
\begin{aligned}
J_{\varphi_l \varphi_p} & =\frac{2 N}{\sigma^2} \operatorname{Re}\left(\alpha_l^* \alpha_p \operatorname{tr}\left(\dot{\mathcal{A}}\left(\varphi_p\right) \mathbf{R}_{sc} \dot{\mathcal{A}}^{\Htran}\left(\varphi_l\right)\right)\right), \\
\mathbf{J}_{{\boldsymbol{\alpha}}_l {\boldsymbol{\alpha}}_p} & =\frac{2 N}{\sigma^2} \operatorname{Re}\left([1, j]^{\Htran}[1, j] \operatorname{tr}\left(\mathcal{A}\left(\varphi_p\right) \mathbf{R}_{sc} \mathcal{A}^{\Htran}\left(\varphi_l\right)\right)\right), \\
\mathbf{J}_{\varphi_l {\boldsymbol{\alpha}}_p} & =\frac{2 N}{\sigma^2} \operatorname{Re}\left(\alpha_l^* \operatorname{tr}\left(\mathcal{A}\left(\varphi_p\right) \mathbf{R}_{sc} \dot{\mathcal{A}}^{\Htran}\left(\varphi_l\right)\right)[1, j]\right),
\end{aligned} \vspace{-2mm}
\end{equation}
where $\dot{\mathcal{A}}(\varphi_i)$ is the derivative of  $\mathcal{A}(\varphi_i)$ with respect to $\varphi_i$, and $i \in \{p,l\}$. $\alpha_l^*$ is the conjugate of $\alpha_l$.
When the number of symbols is sufficiently large, the covariance matrix of the transmit signal, $\vect{R}_{sc}$, can be approximated as 
\vspace{-2mm}
\begin{equation}
\vect{R}_{sc}\approx \sum_{k=1}^{K}\vect{w}_k\vect{w}_k^{\Htran} + \vect{R}_{s}, \vspace{-2mm}
\end{equation}
where $\vect{R}_{s}$ is the covariance matrix of $\sum_{t=1}^{{\bar{T}}}\bar{\vect{w}}_t\bar{s}_t[n]$.
In this letter, we focus only on the characterization of the CRLB for DoA estimation as in \cite{bekkerman2006target, song2022intelligent}, since the target information is harder to extract from $\alpha$ as it both depends on the RCS and the round-trip path loss. The CRLB for the DoA estimate can be given as 
\vspace{-2mm}
\begin{equation}
\mathrm{CRLB}_{\boldsymbol{\varphi}} =  \vect{J}^{-1}_{\boldsymbol{\varphi}} = \left[\vect{J}_{\boldsymbol{\varphi}\boldsymbol{\varphi}}-\vect{J}_{\boldsymbol{\varphi}\boldsymbol{\alpha}}\vect{J}_{\boldsymbol{\alpha}\boldsymbol{\alpha}}^{-1}\vect{J}_{ \boldsymbol{\varphi} \boldsymbol{\alpha}}^{\Ttran}\right]^{-1}.
\label{eq:CRLB}
\vspace{-2mm}
\end{equation}

The diagonal elements of the CRLB matrix can be interpreted as lower bounds on the estimation error variances of the individual targets' DoAs. In the following section, we investigate different sensing options. 

\vspace{-2mm}

\section{Multi-Target ISAC in Massive MIMO}
\vspace{-2mm}

 We consider an ISAC system, where sensing is integrated into the existing communication network. Therefore, the considered ISAC system should guarantee the quality of service requirements of each UE while aiming for the best possible sensing performance for multiple targets.
 
In this letter, we aim to minimize the overall estimation error performance and choose our utility function as the sum of the CRLB values regarding the DoA estimation as $ \tr(\vect{J}^{-1}_{\boldsymbol{\varphi}})$.\footnote{The CRLB is a function of the unknown parameters. In optimizing the CRLB, we utilize rough estimates of the available parameters, which is a typical approach in target tracking scenarios \cite{isac_CRB_main}.} 

\vspace{-4mm}
\subsection{Joint Sensing and Communication Precoding}
Denoting $\vect{R}_k = \vect{w}_k\vect{w}_k^{\Htran}$, 
the problem for the optimal precoding design can be cast as
\vspace{-2mm}
    \begin{subequations}
\begin{align}
  &  \underset{{\vect{R}_1,\ldots,\vect{R}_K,  \vect{R}_{s}}\succeq \vect{0}}{\text{minimize}}  \quad \operatorname{tr}\left(\left[\vect{J}_{\boldsymbol{\varphi}\boldsymbol{\varphi}}-\vect{J}_{\boldsymbol{\varphi}\boldsymbol{\alpha}}\vect{J}_{\boldsymbol{\alpha}\boldsymbol{\alpha}}^{-1}\vect{J}_{ \boldsymbol{\varphi} \boldsymbol{\alpha}}^{\Ttran}\right]^{-1}\right) \label{eq:joint_communication_sensing_precoding_main:objective}\\
  & \text{subject to}  \nonumber \\
      &\frac{\vect{h}_k^{\Ttran}\vect{R}_k\vect{h}_k^*}{\sum_{i=1,i\neq k}^K\vect{h}_k^{\Ttran}\vect{R}_i\vect{h}_k^*+ \vect{h}_k^{\Ttran}\vect{R}_s\vect{h}_k^* + \sigma^2}\geq  \gamma_k^{\rm thr}, \quad \forall k ,\label{eq:joint_communication_sensing_precoding_main:SINR} \\
     & \sum_{k=1}^{K} \operatorname{tr}\left(  \vect{R}_k \right) + \operatorname{tr}\left(  \vect{R}_{s} \right) \leq P_{\mathrm{max}} ,\label{eq:joint_communication_sensing_precoding_main:total_power} \\
    & \vect{R}_k = \vect{w}_k\vect{w}_k^{\Htran}, \quad \forall k. \label{eq:joint_communication_sensing_precoding_main:rank}
\end{align}  
\label{eq:joint_communication_sensing_precoding_main}
\end{subequations}
The objective function \eqref{eq:joint_communication_sensing_precoding_main:objective} is the sum of the CRLBs on the error variance for the DoA estimation of all targets. \eqref{eq:joint_communication_sensing_precoding_main:SINR} is the SINR constraint of UEs, where $2^{\frac{\mathrm{R}_k^{\rm thr}}{B}}-1 = \gamma_k^{\rm thr}$ and $\mathrm{R}_k^{\rm thr}$ is the desired rate of UE $k$. \eqref{eq:joint_communication_sensing_precoding_main:total_power} is the total transmit power constraint. \eqref{eq:joint_communication_sensing_precoding_main:rank} is a non-convex constraint and ensures that UE precoding vectors can be obtained by $\vect{R}_k$. Therefore, the problem needs reformulation to become convex. First, we consider the SDR of the problem by removing \eqref{eq:joint_communication_sensing_precoding_main:rank}.  Later, to tackle the objective function, we define a new symmetric matrix such that  $\tr(\vect{D}^{-1}) \geq \tr(\vect{J}^{-1}_{\boldsymbol{\varphi}})$ under the condition $\vect{J}_{\boldsymbol{\varphi}} \succeq \vect{D} $. By defining this condition as a semi-definite constraint from the Schur complement, the relaxed problem can be equivalently expressed as
\vspace{-2mm}
    \begin{subequations}
\begin{align}
  &  \underset{{\vect{R}_1,\ldots,\vect{R}_K,  \vect{R}_{s}, \vect{D}}\succeq \vect{0}}{\text{minimize}}  \quad \tr\left(\vect{D} ^{-1}\right) \label{eq:joint_communication_sensing_precoding_convex_objective} \\
  &  \,\,\, \text{subject to} \quad \eqref{eq:joint_communication_sensing_precoding_main:SINR}, \eqref{eq:joint_communication_sensing_precoding_main:total_power} \nonumber \\
  &   \begin{bmatrix} \vect{J}_{\boldsymbol{\varphi}\boldsymbol{\varphi}} - \vect{D} & \vect{J}_{\boldsymbol{\varphi}\boldsymbol{\alpha}} \\
    \vect{J}_{\boldsymbol{\varphi}\boldsymbol{\alpha}}^T & \vect{J}_{\boldsymbol{\alpha}\boldsymbol{\alpha}}
    \end{bmatrix} \succeq \vect{0}. \label{eq:joint_communication_sensing_precoding_convex_Schur} 
\end{align}  
\label{eq:joint_communication_sensing_precoding_convex}
\end{subequations}\eqref{eq:joint_communication_sensing_precoding_convex} is a convex problem that can be solved by any convex solver with semi-definite programming. Due to the SDR, the solution of \eqref{eq:joint_communication_sensing_precoding_convex} may not be rank-1, consequently \eqref{eq:joint_communication_sensing_precoding_convex} is not equivalent to \eqref{eq:joint_communication_sensing_precoding_main}. However, we utilized it as a lower bound on the performance for the joint sensing and communication precoding. By using the methods described in \cite{randomization}, a sub-optimal solution can be obtained to the original problem. However, we will not analyze this case since later in numerical analysis, we observed that designing precoding for only the target signals performs very close to the joint precoding lower bound, making the joint design inefficient due to the high complexity.

\vspace{-4mm}
\subsection{Sensing Precoding}

Since Massive MIMO systems have a surplus of antennas, regularized zero-forcing (RZF) is nearly optimal for communications, and the main challenge is power allocation \cite{massivemimobook}.
Hence, instead of joint sensing and communication precoding optimization, we will consider using  RZF for fixed UE precoding while only optimizing the precoding for sensing and the UE power allocation.
This will allow us to reduce the problem size considerably, and we will analyze the corresponding performance loss. 
The unit-norm RZF precoding vector can be expressed as ${\vect{v}}_{i} = \frac{\tilde{\vect{v}}_{i}}{\| \tilde{\vect{v}}_{i}\|}$, where 
\vspace{-2mm}
\begin{equation}
    \tilde{\vect{v}}_{i} = \left[\vect{H}\left(\vect{H}^{\Htran}\vect{H}+ \Omega \vect{I}_{K} \right)^{-1}\right]_{:, i}, \vspace{-2mm}
\end{equation}
and $\vect{H} \in \mathbb{C}^{M \times K}$ is the downlink channel matrix constructed by the columns of the complex conjugates of the UE channel vectors. $\Omega$ denotes the regularization coefficient. In this case, the complete precoding vector for UE $i$ can be obtained by $\vect{w}_i =\sqrt{p_i}\vect{v}_{i}$, where $ p_i$ denotes the allocated power to UE $i$. 

We define the matrix $\vect{A}_i\in \mathbb{R}^{T\times T}$ whose $(l,p)$th entry is given as
\vspace{-2mm}
\begin{equation}
\vect{A}_{i,l,p}  =\frac{2 N}{\sigma^2} \operatorname{Re}\left(\alpha_l^* \alpha_p \operatorname{tr}\left(\dot{\mathcal{A}}\left(\varphi_p\right) \vect{V}_i\dot{\mathcal{A}}^{\Htran}\left(\varphi_l\right)\right)\right), \vspace{-2mm}
\end{equation}
where $\vect{V}_i = \vect{v}_i\vect{v}_i^{\Htran}, i =1,\ldots, K$ and $\vect{V}_{K+1} = \vect{R}_{s}$.
Then we have
$\vect{J}_{\boldsymbol{\varphi}\boldsymbol{\varphi}}=\sum_{i=1}^{K}p_i\vect{A}_i + \vect{A}_{s}$, where $\vect{A}_{s}=\vect{A}_{K+1}$. Similarly, we define $\vect{B}_i\in\mathbb{R}^{T \times 2T}$ whose $(l,2p-1)$th and $(l,2p)$th entries are given as a row vector
\vspace{-2mm}
\begin{equation}
    \vect{B}_{i,l,2p-1:2p} = \frac{2 N}{\sigma^2} \operatorname{Re}\left(\alpha_l^* \operatorname{tr}\left(\mathcal{A}\left(\varphi_p\right) \vect{V}_i \dot{\mathcal{A}}^{\Htran}\left(\varphi_l\right)\right)[1, j]\right).
    \vspace{-2mm}
\end{equation}
Then, we have $\vect{J}_{\boldsymbol{\varphi}\boldsymbol{\alpha}}=\sum_{i=1}^{K}p_i\vect{B}_i + \vect{B}_{s}$,  where $\vect{B}_{s}=\vect{B}_{K+1}$. Finally, we define $\vect{C}_i\in\mathbb{R}^{2T\times 2T}$ whose $(2l-1:2l, 2p-1:2p)$th block is given as
\vspace{-2mm}
\begin{align}
&\vect{C}_{i,2l-1:2l,2p-1:2p} \nonumber\\
&= \frac{2 N}{\sigma^2} \operatorname{Re}\left([1, j]^{\Htran}[1, j] \operatorname{tr}\left(\mathcal{A}\left(\varphi_p\right) \vect{V}_i\mathcal{A}^{\Htran}\left(\varphi_l\right)\right)\right).
\end{align}
Then, it holds that $\vect{J}_{\boldsymbol{\alpha}\boldsymbol{\alpha}}=\sum_{i=1}^{K}p_i\vect{C}_i + \vect{C}_{s}$,  where $\vect{C}_{s}=\vect{C}_{K+1}$. The Fisher information matrix corresponding to the DoA estimates becomes
\vspace{-2mm}
\begin{align}
 \vect{J}_{\boldsymbol{\varphi}} =  &\sum_{i=1}^{K}p_i\vect{A}_i + \vect{A}_s  -\left(\sum_{i=1}^{K}p_i\vect{B}_i + \vect{B}_s\right) \times \nonumber \\ &\left(\sum_{j=1}^{K}p_j\vect{C}_j+ \vect{C}_s\right)^{-1} \left(\sum_{r=1}^{K}p_r\vect{B}_r^{\Ttran}+ \vect{B}^{\Ttran}_s\right).
\end{align}

Since we remove the UE precoding vectors from the problem, SDR is not required, and the problem can be expressed in convex form by only using the Schur complement.  The problem formulation for the optimal power control and sensing precoding design can be expressed as
\vspace{-2mm}
    \begin{subequations}
\begin{align}
  &  \underset{{{p}_1,\ldots,{p}_K\geq 0,  \vect{R}_s, \vect{D}\succeq \vect{0}}}{\text{minimize}}  \quad \tr\left(\vect{D} ^{-1}\right) \label{eq:only_sensing_precoding:objective} \\
  &  \text{subject to} \nonumber \\ & \frac{p_k|\vect{h}_k^{\Ttran}\vect{v}_k|^2}{\sum_{i=1,i\neq k}^Kp_i|\vect{h}_k^{\Ttran}\vect{v}_i|^2+ \vect{h}_k^{\Ttran}\vect{R}_{s}\vect{h}_k^* + \sigma^2}\geq \gamma_k^{\rm thr},  \forall k , \label{eq:only_sensing_precoding:SINR}  \\
  &   \begin{bmatrix} \sum_{i=1}^{K}p_i\vect{A}_i + \vect{A}_s - \vect{D} & \sum_{i=1}^{K}p_i\vect{B}_i + \vect{B}_s \\
\sum_{i=1}^{K}p_i\vect{B}^{\Ttran}_i + \vect{B}^{\Ttran}_s & \sum_{i=1}^{K}p_i\vect{C}_i+ \vect{C}_s
    \end{bmatrix} \succeq \vect{0} , \label{eq:only_sensing_precoding:Schur}  \\
     & \sum_{i=1}^{K} p_i + \operatorname{tr}\left(  \vect{R}_s \right) \leq P_{\rm max}. \label{eq:only_sensing_precoding:sum_power}
\end{align}  
\label{eq:only_sensing_precoding}
\end{subequations}
\eqref{eq:only_sensing_precoding} can be solved with any convex programming solver.

\vspace{-4mm}

\subsection{Orthogonal Sensing and Communication}
An alternative way of operation is to transmit communication and sensing signals using orthogonal resources. In this way, the interference between the communication and sensing signals will be eliminated, but at the cost of reducing the number of channel uses utilized for each type of transmission. To guarantee the same rate as the previous methods, UEs will be required to have a higher SINR constraint. We let $0<\eta<1$ to show the proportion of communication duration.  The SINR constraint becomes $\gamma^{\mathrm{thr, ort}}_k = (\gamma^{\mathrm{thr}}_k+1)^{1/\eta}-1$ to guarantee the same rate constraints as with the joint sensing and communication.  Similarly, the estimation error variance will be increased proportionally to $1/(1-\eta)$. The problem for orthogonal sensing and communication can be stated as 
\vspace{-2mm}
    \begin{subequations}
\begin{align}
  &  \underset{{{p}_1,\ldots,{p}_K\geq 0,  \vect{R}_s, \vect{D}\succeq \vect{0}}}{\text{minimize}}  \quad \tr\left(\vect{D} ^{-1}\right) \label{eq:orthogonal_sensing_precoding:objective} \\
  &  \text{subject to} \quad  \nonumber \\ & \frac{p_k|\vect{h}_k^{\Ttran}\vect{v}_k|^2}{\sum_{i=1,i\neq k}^Kp_i|\vect{h}_k^{\Ttran}\vect{v}_i|^2 + \sigma^2}\geq \gamma^{\mathrm{thr, ort}}_k ,  \forall k ,\label{eq:orthogonal_sensing_precoding:SINR}  \\
  &   \begin{bmatrix}  \vect{A}_s - \vect{D} &  \vect{B}_s \\
 \vect{B}^{\Ttran}_s &  \vect{C}_s
    \end{bmatrix} \succeq \vect{0}.\label{eq:orthogonal_sensing_precoding:Schur} \\
& \eta \sum_{i=1}^{K} p_i + (1-\eta)\operatorname{tr}\left(  \vect{R}_s \right) \leq P_{\rm max}.   \label{eq:orthogonal_sensing_precoding:power_limit} 
\end{align}  
\label{eq:orthogonal_sensing_precoding} 
\end{subequations}
Note that in both \eqref{eq:orthogonal_sensing_precoding:SINR} and  \eqref{eq:orthogonal_sensing_precoding:Schur}, the effect of interference is removed. This problem can also be solved by conventional convex programming solvers. The structure of the optimal sensing precoder is influenced by non-linear CRLB terms, leading it to lie primarily in the subspace of both the derivative of the target channels and nullspace of the user channels, thereby optimizing sensing performance while minimizing interference.
\vspace{-3mm}
\subsection{Power Allocation}
Another possible way to simplify the problem even further is to assign individual unit-norm precoding vectors to the sensing signals and only optimize the power allocation variables. Similar to the communication UEs, we can construct the sensing precoders $\bar{\vect{w}}_t = \sqrt{\bar{p}_t}\bar{\vect{v}}_t$, where only $\bar{p}_t$  is optimized along with the UE power allocation variables, where we set $\bar{T}=T$. One good way to select the unit-norm precoding vector, as in \cite{buzzi_target_precoding}, $\bar{\vect{v}}_t$, is to choose a zero-forcing (ZF) precoding, by projecting the target channel onto the nullspace spanned by the user channels. We let $\hat{\vect{U}} = \left[\vect{h}_1, \ldots, \vect{h}_K \right] $, and 
\vspace{-2mm}
\begin{equation}
    \tilde{\vect{v}}_t = (\vect{I}_{M} - \hat{\vect{U}}\hat{\vect{U}}^{\Htran})  \vect{a}(\varphi_t), \vspace{-2mm}
\end{equation}
and $\bar{\vect{v}}_t = \tilde{\vect{v}}_t/ \|\tilde{\vect{v}}_t \|$.
In this way, the interference caused by the sensing signal to communication will be limited. We can reformulate the sensing related parameters as $\vect{J}_{\boldsymbol{\varphi}\boldsymbol{\varphi}}=\sum_{i=1}^{K}p_i\vect{A}_i+\sum_{t=1}^T\bar{p}_t\bar{\vect{A}}_t$, $\vect{J}_{\boldsymbol{\varphi}\boldsymbol{\alpha}}=\sum_{i=1}^{K}p_i\vect{B}_i+\sum_{t=1}^T\bar{p}_t\bar{\vect{B}}_t$, and $\vect{J}_{\boldsymbol{\alpha}\boldsymbol{\alpha}}=\sum_{i=1}^{K}p_i\vect{C}_i+\sum_{t=1}^T\bar{p}_t\bar{\vect{C}}_t$, where $\bar{\vect{A}}_t$, $\bar{\vect{B}}_t$, and $\bar{\vect{C}}_t$, for $t=1,\ldots,T$ are defined in a manner similar to that previously described.
The problem can be equivalently expressed as
    \begin{subequations}
\begin{align}
  &  \underset{{p_1,\ldots,p_{K},\bar{p}_1,\ldots,\bar{p}_T\geq 0, \vect{D}\succeq \vect{0}}}{\text{minimize}}  \quad \tr\left(\vect{D} ^{-1}\right) \\
  &  \text{subject to} \nonumber \\ 
  & \frac{p_k|\vect{h}_k^{\Ttran}\vect{v}_k|^2}{\sum\limits_{i=1,i\neq k}^Kp_i|\vect{h}_k^{\Ttran}\vect{v}_i|^2+ \sum\limits_{t=1}^T\bar{p}_t|\vect{h}_k^{\Ttran}\bar{\vect{v}}_t|^2 +
 \sigma^2}\geq \gamma_k^{\rm thr},  \forall k \label{eq:SINR_power_allocation} \\
  &   \begin{bmatrix} \sum_{i=1}^{K}p_i \vect{A}_i+\sum_{t=1}^T\bar{p}_t\bar{\vect{A}}_t - \vect{D} & \sum_{i=1}^{K}p_i\vect{B}_i+\sum_{t=1}^T\bar{p}_t\bar{\vect{B}}_t \\
    \sum_{i=1}^{K}p_i\vect{B}_i^{\Ttran}+\sum_{t=1}^T\bar{p}_t\bar{\vect{B}}_t^{\Ttran} & \sum_{i=1}^{K}p_i\vect{C}_i+\sum_{t=1}^T\bar{p}_t\bar{\vect{C}}_t
    \end{bmatrix} \nonumber\\
    &\succeq \vect{0} ,\label{eq:power_allocation:constraint_Schur_1} \\
  & \sum_{i=1}^{K+T} p_i  \leq P_{\mathrm{max}}.
\end{align}  
\label{eq:power_allocation}
\end{subequations}
{\eqref{eq:power_allocation} can also be solved with conventional convex programming solvers considering any linear precoders for communication and sensing targets.  Note that, the chosen unit-norm precoding vectors for the sensing targets are capable of canceling the interference of the sensing signal to the UEs. Consequently, the interference term in \eqref{eq:SINR_power_allocation} becomes zero, $\sum_{t=1}^T\bar{p}_t|\vect{h}_k^{\Ttran}\bar{\vect{v}}_t|^2= 0$. } Compared to the previously presented algorithms, power allocation has the smallest problem size and, therefore, requires the least amount of computation.

\begin{figure*}[tb]
	\begin{center}	
		\subfigure[]{
			\label{fig:M32_T}
			\includegraphics[width=0.34\linewidth]{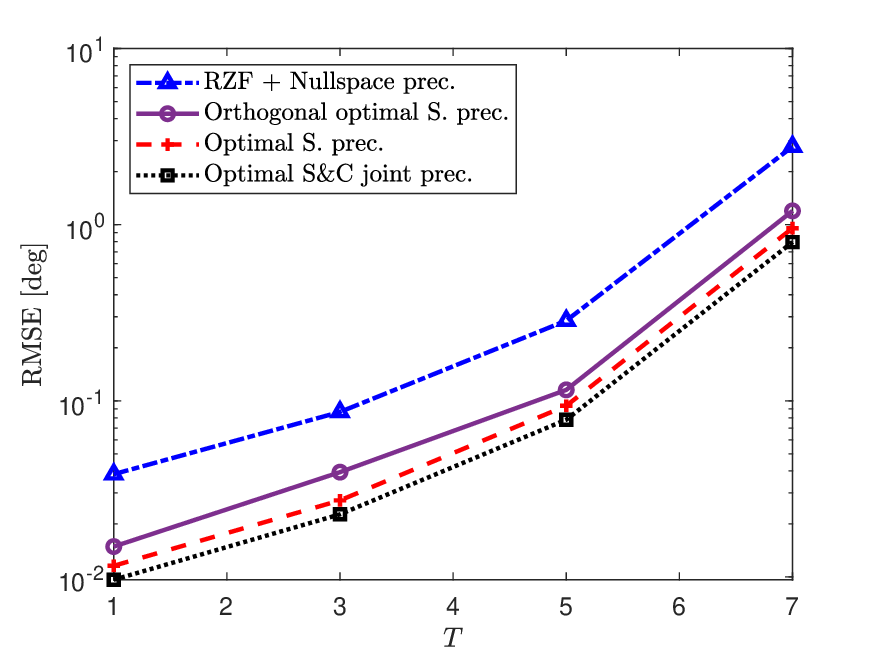} \vspace{-10mm}
 }
   \hspace{-8mm}
		\subfigure[]{
			\label{fig:M64_T}
		\includegraphics[width=0.34\linewidth]{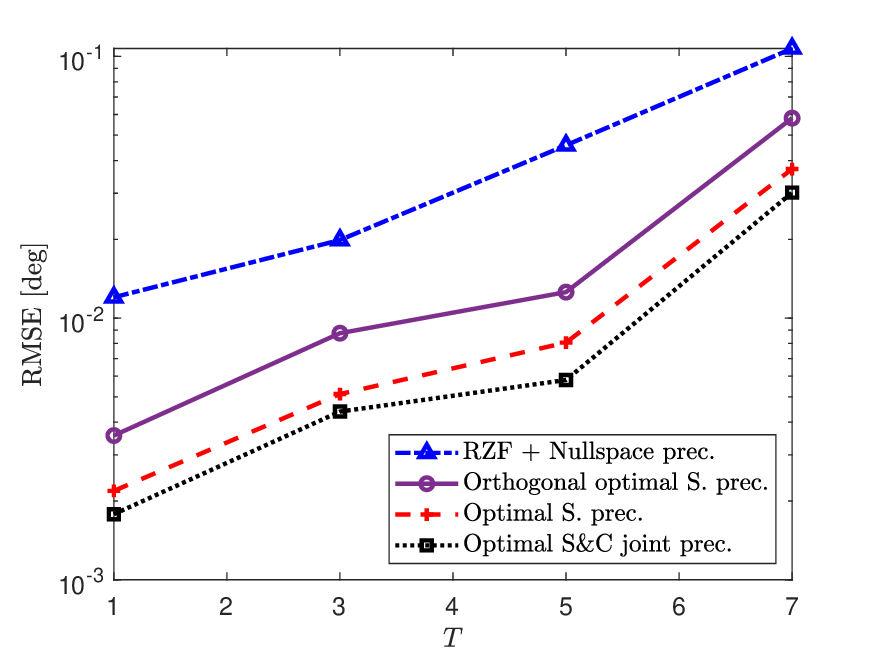} \vspace{-10mm}
 }
    \hspace{-8mm}
		\subfigure[]{
			\label{fig:orthogonal_vs_joint}
		\includegraphics[width=0.34\linewidth]{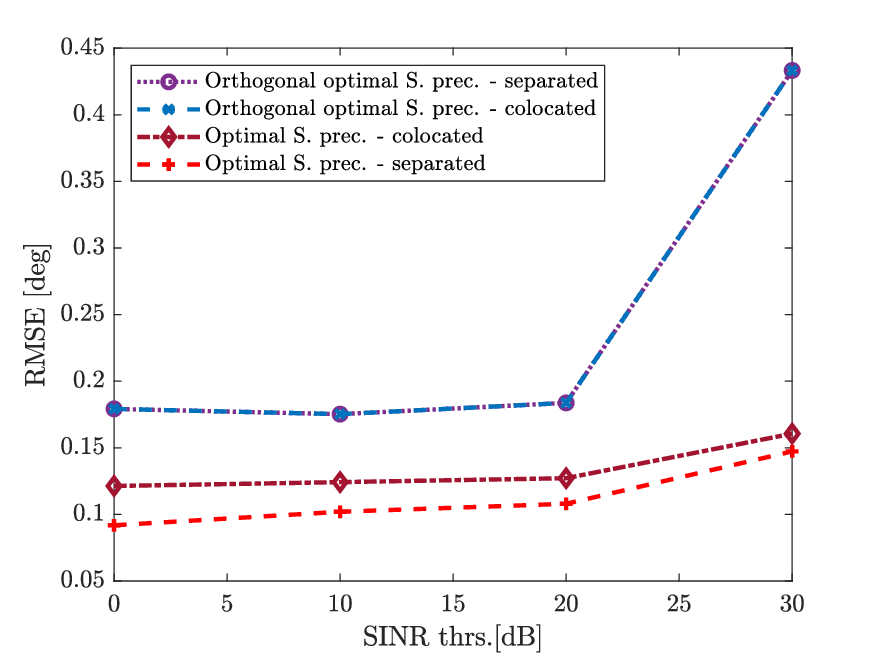} \vspace{-10mm}
 }
  
	\end{center}
	\vspace{-5mm}
	\caption{Performance comparison of the RMSE achieved by different algorithms with varying numbers of targets for (a) $M=32$, (b) $M=64$. RZF+nullspace precoding, orthogonal optimal sensing precoding, optimal sensing precoding, optimal sensing and communication joint precoding correspond to the solutions of  \eqref{eq:power_allocation}, \eqref{eq:orthogonal_sensing_precoding}, \eqref{eq:only_sensing_precoding}, and \eqref{eq:joint_communication_sensing_precoding_convex}, respectively. (c) RMSE performance of orthogonal and joint sensing and communication when UEs and targets are co-located or separated.}
	\label{fig:T_effect}
 \vspace{-2mm}
\end{figure*}
 \vspace{-2mm}

\section{Numerical Results}
We consider a simulation area of  $500\text{\,m} \times 500\text{\,m}$. The UEs are randomly generated within this area. We investigate two scenarios for target locations: deterministic and random. In the deterministic target location scenario, we assume that the distance between the target and the BS is $150\text{\,m}$, and UEs are randomly distributed over the entire $500\text{\,m} \times 500\text{\,m}$ area. In the random target location scenario, both UEs and targets are randomly generated within neighborhood regions of $75\text{\,m} \times 75\text{\,m}$, with their centers located at the Cartesian coordinates $[140, -100]$ and $[170, 30]$, respectively. For both scenarios, we generate $200$ random instances. The root-mean-square error (RMSE) represents the square root of the average CRLB in degrees.  The maximum total transmit power is set to $10$\,W. We set $M=64$, $N=100$, and $K=8$ unless otherwise stated \footnote{ We observe that $K$ does not have a significant effect except RZF+nullspace precoding, where the performance of it improves as the number of UEs increases.}. The path loss for the UE channels is modeled by the 3GPP Urban Microcell model \cite{3GPP}, and a correlated Rayleigh fading channel is assumed. The carrier frequency is $1.9\text{\,GHz}$ and the bandwidth is $20$\,MHz.  The path loss for the target channels is modeled by the two-way radar range equation \cite{zinat_globecomm}.  RCS is modeled by the Swerling-I model where $\alpha \sim \mathcal{CN}(0,1)$. $\gamma^{\mathrm{thr}}_k$ is set to 10\,dB in the simulations. CVX with MATLAB is utilized to solve the given problems. 

Fig.~\ref{fig:T_effect} shows the sensing performance of the proposed algorithms for different numbers of targets. We consider a random target generation scenario. According to the figure, precoding optimization for sensing purposes is essential since only optimizing the power does not provide efficient performance.
However, if we use fixed RZF precoding and optimize the sensing precoding, we obtain performance very close to joint S\&C precoding lower bound in both the $M=32$ and $M=64$ cases in Figs.~\ref{fig:M32_T} and \ref{fig:M64_T}, respectively.
  
Another notable result is that performing sensing and communication orthogonally results in inferior performance compared to the joint transmission of these signals. This trend is more pronounced in Fig.~\ref{fig:M64_T} due to the higher interference cancellation ability and sharper beams with the increased number of antennas. Due to the close performance of sensing precoding and joint S\&C precoding lower bound, we did not investigate the real achievable performance of joint S\&C precoding, but that can be obtained by using the methods given in \cite{randomization}. The effect of $N$ can be seen by factoring it in \eqref{eq:CRLB}, where $N$ is inversely proportional to the CRLB, meaning a tenfold increase in $N$ improves sensing performance by a factor of 10.

To analyze the trade-off between the sensing and the communication, we consider two different UE and target location scenarios in Fig.~\ref{fig:orthogonal_vs_joint}.  In the first scenario, targets and UEs are separated, i.e., they are randomly generated in different regions as described above.  In the second scenario, targets and UEs are co-located, i.e., they are randomly generated within a square region of $75\text{\,m} \times 75\text{\,m}$ centered at  $[170, 30]$ \footnote{In both scenarios, the angles between any  UEs or any targets are not allowed to be less than $0.1$ degrees to eliminate outlier cases.}. We set the RCS variance at $-20$\,dB to consider a challenging sensing task for both scenarios.   We compare the performance of two different methods: (i) time sharing between communication and sensing, and (ii) orthogonal time division of sensing and communication signals. In both cases, we consider RZF precoding for the communication signals, and we optimize the sensing precoding. The corresponding problems for time division and time sharing are  \eqref{eq:orthogonal_sensing_precoding} and \eqref{eq:only_sensing_precoding}, respectively. Orthogonal time sharing performs similarly for UE and target separation or co-location scenarios since communication and sensing signals do not interfere in this case, and UEs and targets have approximately the same distance to the BS in both scenarios. We observe that time sharing performs well compared to the orthogonal time allocation. This demonstrates that the proposed sensing precoding design can benefit from the communication signals even if the communication signals are not optimized for the sensing. The power assigned for communication helps the sensing, while the interference created by sensing to UEs is also optimized to satisfy the SINR constraints of the UEs. Conversely, in the orthogonal time sharing, time resources are shared between sensing and communication, dropping the time duration used for sensing to $50\%$. Consequently, as the SINR constraints for UEs become more stringent, more power is allocated to communication, and the reduction in sensing performance becomes more prominent.

Table \ref{tab:time_algorithm} provides the algorithms' average run time for different numbers of targets. The cost of joint S\&C precoding is considerably higher (more than $100$ times higher than power allocation) than the low-complexity algorithms provided. Due to the rank reduction algorithms, joint S\&C precoding would require much more time than the values given in the Table. Overall, we observe that sensing precoding optimization provides performance close to the lower bound while requiring much less complexity than joint S\&C precoding optimization.

\begin{table}[tb]
\vspace{-7mm}
\caption{Average run time of the proposed algorithms in seconds considering $M=64$.}
\vspace{-2mm}
\centering
\begin{tabular}{|l|l|l|l|}
\hline T & Power allocation & Sensing prec. & Joint S\&C prec.  \\ \hline
1 & 1.6             & 7.0                   & 68.5           \\
3 & 2.9             & 30.8                  & 511.6            \\
5 & 4.1             & 74.1                  & 1382.8          \\
7 & 4.1             & 135.9                 & 2620.0 \\ \hline         
\end{tabular}
\label{tab:time_algorithm}
\vspace{-5mm}
\end{table}

\vspace{-4mm}
\section{Conclusions}
In this letter, we have studied monostatic multi-target sensing in a downlink massive MIMO system. Four ISAC designs have been considered: jointly optimized communication and sensing precoding, optimized sensing precoding and communication power allocation, orthogonal sensing and communications, and optimized power allocation for both communication and sensing. We transformed each design into a convex optimization problem with semi-definite constraints using the Schur complement. The numerical analysis demonstrates that optimized sensing precoding with heuristic RZF precoding for UEs performs similarly to the joint S\&C precoding lower bound but has roughly $20$ times less computational complexity. The orthogonal assignment of communication and sensing tasks performs poorly, indicating the benefits of non-orthogonal resource sharing for sensing and communication. As a future work, dynamic joint optimization of sensing duration and precoding based on the changing target conditions and user channels can potentially improve the performance of orthogonal communication and sensing.

\vspace{-5mm}
\bibliographystyle{IEEEtran}
\bibliography{IEEEabrv,refs}

\end{document}